\documentclass[aps,showpacs]{revtex4}
\usepackage{amssymb}
\usepackage{graphicx}

\begin{document}

\title{The Kondo-lattice state in the presence of Van Hove singularities:
a next-leading order scaling description}
\author{V. Yu. Irkhin}
\affiliation{M. N. Mikheev Institute of Metal Physics, 620990 Ekaterinburg, Russia}
\email{Valentin.Irkhin@imp.uran.ru}

\begin{abstract}
A renormalization group treatment of the Kondo model with a logarithmic Van
Hove singularity in the electron density of states is performed within
next-leading order scaling, different magnetic phases being considered.
The effective coupling constant remains small, renormalized magnetic moment
and spin-fluctuation frequency decreasing by several orders of magnitude.
Thus wide non-Fermi-liquid behavior regions are found from the scaling
trajectories in a broad interval of the bare coupling parameter.
Applications to physics of itinerant magnetism  are discussed.
\end{abstract}

\maketitle


\section{Introduction}
Anomalous rare-earth and actinide compounds
(Kondo lattices and heavy-fermion systems) are studied extensively starting
from the middle of 1980s \cite{Stewart}.  Besides heavy-fermion
features (huge electronic specific heat), they exhibit the
non-Fermi-liquid (NFL) behavior: logarithmic or anomalous power-law
temperature dependences of magnetic susceptibility and specific heat \cite{Stewart1l}. The magnetism of these systems is also highly interesting and has both localized and
itinerant features; in particular, magnetic moment can be strongly reduced or unstable \cite{kondo,Coleman1,I17}. Thus a combined treatment of Kondo and
itinerant-electron systems becomes usual \cite{Ohkawa,Vojta}. NFL behavior and strongly enhanced electronic specific heat are observed also in some d-systems \cite{Stewart1l}, including nearly ferromagnetic ruthenates \cite{rut,rut1,rut2}.

Main role in the physics of the Kondo lattices belongs to  interplay of
 on-site Kondo screening of magnetic moments and intersite exchange
interactions inducing magnetic order. This idea was developed in a series of
the papers \cite{IKZ1,kondo} treating the mutual renormalization of two
energy scales: the Kondo temperature $T_{K}$ and
characteristic spin-fluctuation frequency $\overline{\omega }$. The
corresponding scaling consideration of this renormalization process in the $%
s-f$ exchange model \cite{kondo,731} yields, depending on the values of bare
parameters, both the \textquotedblleft usual\textquotedblright\ states (a
non-magnetic Kondo lattice or a magnetic state with weak Kondo corrections)
and the peculiar magnetic Kondo-lattice state, including the NFL behavior.
However, the region of the latter behavior depends strongly on the approximations and
 concrete models of electron and magnetic structure \cite{I11,I16,731}.
The NFL behavior can be related to peculiar features of electron and spin
fluctuation spectrum, as well as in the case of itinerant systems (which are usually
described by the Hubbard model).

It is well known that peculiarities of bare electron structure play an important role in the formation of magnetism. The case of singular
density of states (DOS) for the Kondo lattices was considered in Ref.\cite{I11} within the lowest-order scaling. Although a considerable increase of NFL region was obtained,  the situation did not qualitatively change in comparison with the smooth DOS case.

At the same time, the NFL behavior occurs naturally in the one-impurity $M$%
-channel Kondo model \cite{Col,Gan,Cox}. This model, which assumes existence
of degenerate electron bands, explains power-law or logarithmic behavior of
electronic specific heat and magnetic susceptibility \cite{Cox}.
Physically, such a behavior is connected with overscreening of impurity spin
by conduction electrons in many channels. The model permits a consistent
scaling investigation in the next-leading approximation. A characteristic
feature of this approximation is occurrence of an intermediate fixed point.
This is reasonable for $M>2$ since the fixed point is within the
weak-coupling region (however, the marginal case $M=2$ requires a more
accurate consideration). On the other hand, for $M=1$ the fixed point is
unphysical. The situation for the lattice is more complicated, especially in
the case of singular DOS, since the singularities change the structure of
perturbation theory.


In the present paper we treat the Kondo lattice model with the electron
spectrum containing a logarithmic DOS singularity to
compare the results of leading and next-leading order scaling. Such a
singularity is typical, in particular, for the two-dimensional case. The
scaling equations are discussed in Sect.2. The numerical results are
presented in Sect.3  for the cases of a paramagnet with smooth spin
spectral function and a ferromagnet (singular spin spectral function).

\section{Scaling equations}

We use the degenerate-band (multichannel) Kondo lattice model
\begin{equation}
H=\sum_{\mathbf{k}m\sigma }t_{\mathbf{k}}c_{\mathbf{k}m\sigma }^{\dagger }c_{%
\mathbf{k}m\sigma }-I\sum_{im\sigma \sigma ^{\prime }}\mathbf{S}_{i}%
\mbox {\boldmath
$\sigma $}_{\sigma \sigma ^{\prime }}c_{im\sigma }^{\dagger }c_{im\sigma
^{\prime }}^{{}}+H_{f}  \label{1}
\end{equation}%
where $t_{\mathbf{k}}$ is the band energy, $\mathbf{S}_{i}$ are spin-1/2
operators, $I$ is the $s-f$ exchange parameter, $\sigma $ are the Pauli
matrices, $m=1...M$ is the orbital degeneracy index. For the sake of
convenient constructing perturbation theory, we explicitly include the
Heisenberg $f-f$ exchange interaction
\begin{equation}
H_{f}=\sum_{ij}J_{ij}\mathbf{S}_i\mathbf{S}_j
 \label{eq:F.1}
\end{equation}
in the Hamiltonian, although
in fact this interaction is usually the indirect RKKY coupling.
This interaction competes with the Kondo effect and results in occurrence of cutoffs for the corresponding infrared divergences.
Depending on character of $f-f$ interactions, we can treat paramagnetic and various magnetically ordered (ferro- or antiferromagnetic) phases with reduced moments.

The density of states corresponding to the spectrum $t_{\mathbf{k}}$ is
supposed to contain a Van Hove singularity (VHS) near the Fermi level. The simplest example is the square lattice with the spectrum $t_{\mathbf{k}}=2t(\cos
k_{x}+\cos k_{y})$ where we have the density of states%
\begin{eqnarray}
\rho (E) &=&\frac{2}{\pi ^{2}D}K\left( \sqrt{1-\frac{E^{2}}{D^{2}}}\right)
\nonumber \\
&\simeq &\rho F(E),\ F(E)=\ln \frac{D}{|E|}+2\ln 2,~\rho =\frac{2}{\pi ^{2}D}
\end{eqnarray}%
where $K(E)$ is the complete elliptic integral of the first kind, the
bandwidth is determined by $|E|<D=4|t|$. However, considerable singularities can occur for three-dimensional lattices too \cite{Vons}.

At $M>2$ the fixed point lies in a weak coupling region, which makes possible
successful application of perturbation theory and renormalization group approaches.
We apply the \textquotedblleft poor man scaling\textquotedblright\ approach
\cite{And}. This considers the dependence of effective (renormalized) model
parameters (effective $s-f$ coupling and spin-fluctuation frequencies) on
the flow cutoff parameter $C\rightarrow -0.$


To find the equation for the renormalized  coupling parameter $I_{ef}(C)$ we
pick out in the sums for the Kondo corrections the contribution of
intermediate electron states near the Fermi level with $C<t_{\mathbf{k+q}%
}<C+\delta C.$ Bearing in mind a NFL-type behavior, we write down the scaling equations to next-leading order by taking into account the corrections of order of $MI^{3 }$ (see details in Refs. \cite{kondo,I11,I16}):
\begin{equation}
\delta I_{ef}(C)=2\rho I^{2}[F(C)+I\rho MF(C/2)F(-C/2)]\eta (-\frac{%
\overline{\omega }}{C})\delta C/C  \label{ief}
\end{equation}%
where $\overline{\omega }$ is a characteristic spin-fluctuation energy (the ratio $\bar{\omega}/|C|$ is initially assumed to be small within perturbation theory, but can become arbitrary depending on scaling behavior during renormalization process), $\eta (x)$ is a scaling function taking into account spin dynamics and satisfying the condition $\eta (0)=1.$
Strictly speaking, the next-to-leading order contribution is exact in the large-$M$ limit only, but the finite-$M$ case can be treated in agreement with the one-impurity limit (see Sect.3).

For the paramagnetic (PM), ferromagnetic (FM) and antiferromagnetic (AFM) phases  we have
\begin{equation}
\eta ^{PM}(\frac{\overline{\omega }}{C})=\mathrm{{Re}\int_{-\infty }^{\infty
}d\omega \langle \mathcal{J}_{\mathbf{k-k}^{\prime }}(\omega )\rangle
_{t_{k}=t_{k^{\prime }}=E_{F}}\frac{1}{1-(\omega +i0)/C}}
\end{equation}%
\begin{equation}
\eta^{FM,AFM} \left( \overline{\omega }_{ef}/|C|,\delta \right) =\mathrm{{Re}%
\left\langle \left( 1-(\omega _{\mathbf{k-k}^{\prime }}^{{}}+i\delta
)^{2}/C^{2}\right) ^{-1}\right\rangle _{t_{k}=t_{k^{\prime }}=E_{F}}}
\label{eta11}
\end{equation}%
Here $\omega _{\mathbf{q}}$ is the magnon frequency, angle brackets stand
for the average over the wavevector $\mathbf{k}$ on the Fermi surface,
\begin{equation}
\langle A({\mathbf{k}}) \rangle _{t_{\mathbf{k}}=E_{F}} = \sum_{\mathbf{k}}
A({\mathbf{k}} ) \delta(t_{\mathbf{k}} - E_F)  \nonumber
\end{equation}
$\mathcal{J}_{\mathbf{q}}(\omega )$ is the spectral density of the spin
Green's function for the Hamiltonian $H_{f},$ which is normalized to unity, $%
\delta $ is a cutoff owing to damping.

In the simple spin-diffusion approximation for a paramagnet and spin-wave
approximation for the magnetic phases we have \cite{kondo,I16}%
\begin{eqnarray}
\eta ^{PM}(x) &=&\arctan x/x \\
\eta ^{FM}(x) &=&\frac{1}{4x}\ln \{[(1+x)^{2}+\delta ^{2}]/[(1-x)^{2}+\delta
^{2}]\} \\
\eta ^{AFM}(x) &=&-(2x^{2})^{-1}\ln [(1-x^{2})^{2}+4\delta ^{2}]
\label{damp}
\end{eqnarray}%
The scaling functions for the ordered phases contain Van Hove singularities
at $x=1.$

The renormalizations of magnetic moment and spin fluctuation frequency are also obtained from perturbation theory and
are given by \cite{kondo,I11,I16}%
\begin{equation}
\delta \overline{\omega }_{ef}(C)/\overline{\omega }=a\delta \overline{S}%
_{ef}(C)/S=2a\rho ^{2}I^{2}F(C/2)F(-C/2)\eta (-\frac{\overline{\omega }}{C}%
)\delta C/C  \label{aa}
\end{equation}%
The latter result holds for all magnetic phases with $a=1-\alpha $ for the
paramagnetic (PM) phase, $a=2(1-\alpha )$ for the ferromagnetic (FM) phase,
$a=1-\alpha ^{\prime }$ for the antiferromagnetic (AFM) phase. Here $\alpha$ and $\alpha ^{\prime }$ are some averages over the
Fermi surface (see Ref.\cite{kondo}). In the approximation of
nearest neighbors at the distance $ |\mathbf{R|}=d$ one obtains
\begin{equation}
\alpha =\left\vert \langle \exp (i\mathbf{kR}_{2})\rangle _{t_{\mathbf{k}%
}=E_{F}}\right\vert ^{2}\simeq \left( \frac{\sin k_{F}d}{k_{F}d}\right)
^{2}  \label{Alpqf}
\end{equation}%
For the staggered AFM ordering we have
\begin{equation}
\alpha ^{\prime }\simeq b\frac{J_{2}}{J_{1}}\left\vert \left\langle \exp (i\mathbf{kR}_{2})\right\rangle _{t_{\mathbf{k}}=E_{F}}\right\vert ^{2}
\end{equation}%
where $b=2$ and $b=4$ for the square and simple cubic lattices, $J_{1}$ and $J_{2}$ are the
Heisenberg exchange integrals between nearest and next-nearest neighbors ($|J_{1}|\gg
|J_{2}|$) in $H_f$, $\mathbf{R}_{2}$ runs over the next-nearest neighbors. Although $\alpha ^{\prime }=0$ in the nearest-neighbor approximation, this parameter enters physical properties in the NFL regime \cite{731}.

Defining the renormalized and
bare dimensionless coupling constants%
\begin{equation}
g_{ef}(C)=-2\varrho I_{ef}(C),\ g=-2I\varrho
\end{equation}%
and the function
\begin{equation}
 \psi (\xi ) =\ln(\overline{\omega}/\overline{\omega }_{ef}(\xi))
\end{equation}
which determines renormalization of spin dynamics we obtain%
\begin{equation}
\partial g_{ef}(\xi )/\partial \xi =[\xi -\gamma (\xi +\ln 2)^{2}g_{ef}(\xi
)]g_{ef}^{2}(\xi )\Psi (\lambda +\psi -\xi ),~  \label{sc2}
\end{equation}%
\begin{equation}
\partial \psi (\xi )/\partial \xi =a\gamma g_{ef}^{2}(\xi )(\xi +\ln
2)^{2}\Psi (\lambda +\psi -\xi )  \label{sc3}
\end{equation}%
where $\gamma =M/2,$ we put in spirit of scaling consideration $\xi =\ln
|D/C|+2\ln 2\simeq \ln |D/C|$ (note that a constant DOS contribution is
absorbed by the replacement $\xi \rightarrow \xi +c$),
\[
\Psi (\xi )=\eta (e^{-\xi }),\quad \lambda =\ln (D/\overline{\omega })\gg 1.
\]

First we discuss briefly the one-impurity case ($\Psi =1$). The solution of the lowest-order
(one-loop) scaling equation according to (\ref{ief}) yields
\begin{equation}
1/g_{ef}(\xi) =1/g- \xi^2/2
 \label{perturb11}
\end{equation}%
The divergence of $g_{ef}(C)$ occurs at the Kondo temperature%
\begin{equation}
T_{K}\varpropto D\exp \left[ -\left| \frac{\pi ^{2}D}{2I}\right| ^{1/2}%
\right]  \label{perturb}
\end{equation}%
This result  is in agreement with perturbation theory and numerical renormalization group (NRG) results for the singular DOS case, unlike the parquet approach of Ref.\cite{Gogolin} (see  discussion in Refs.\cite{Zhur2008,I11}).

For comparison with the standard Kondo problem, it is instructive to introduce the function
\[
G_{ef}(\xi )=g_{ef}(\xi )\xi.
\]%
Owing to the structure of perturbation theory, this quantity is an effective coupling parameter in the singular DOS situation. In particular, $G_{ef}(\xi )$  (with the replacement $|C|\rightarrow T$) enters  corrections for electronic properties like magnetic susceptibility and specific heat, cf. Ref.\cite{I11}.
When neglecting $\ln 2$ in comparison with $\xi$, the scaling equation for this function takes the form
\begin{equation}
\partial G_{ef}(\xi )/\partial \xi = G_{ef}(\xi )/\xi +[1 -\gamma
G_{ef}(\xi )]G_{ef}^{2}(\xi )
\label{GG}
\end{equation}%
Apart from the first term (which is small  at large $\xi$, i.e. at low energies), the right-hand side does not depend explicitly  on $\xi$. Thus Eq.(\ref{GG}) has the structure of a standard Gell-Mann-Low equation and is similar to the scaling equation for $g_{ef}(\xi)$ in the smooth DOS case. In the latter case the two-loop
equation corresponding to (\ref{sc2}) gives a finite fixed point $g_{ef}(\xi
\rightarrow \infty )=2/M$. It is known that this point is unphysical
(unreachable) for $M=1$, but for $M>2$ the scaling consideration gives a
qualitatively correct description \cite{Cox} (the case $M=2$ is marginal, so
that additional logarithmic factors occur in the physical properties).




Unlike the smooth DOS case, the equation (\ref{sc2}) cannot be solved
analytically even in the one-impurity case, but an asymptotic solution
at large $\xi $ can be obtained:
\begin{equation}
g_{ef}(\xi )=\frac{2}{M}\frac{1}{\xi }+\left( 1-\frac{4\ln 2}{M}\right)
\frac{1}{\xi ^{2}}  \label{expansion}
\end{equation}%
The second term in brackets can change sign, being positive for large $M$ and
negative for small $M$,
so that occurrence of a maximum in the dependence of $g_{ef}(\xi )$ is
possible. Besides that, the factor in (\ref{expansion}) is well determined
only within the $1/M$-expansion. Thus the solution is rather sensitive to
details of approximations.

The above results demonstrate existence of the \textquotedblleft fixed
point\textquotedblright, which is similar to the fixed point in the flat-band case,
\begin{equation}
G_{ef}(\xi )=G^{\ast }=2/M\label{expansion11}
\end{equation}%
The scaling trajectories approach this according the law
\begin{equation}
G^{\ast }-G_{ef}(C)\varpropto 1/\ln |C|,
\end{equation}%
unlike the power law in the case of smooth DOS \cite{Gan,I16}. Note that corresponding $1/\ln T$-dependences are obtained in NRG calculations of impurity magnetic susceptibility and specific heat \cite{Zhur2008,Zhur2016}.

Writing down the Kondo correction to magnetic susceptibility by analogy with Refs.\cite{Gan,kondo,I11},
 we obtain the scaling equation for the effective magnetic moment%
\begin{equation}
\partial \ln S_{ef}(\xi )/\partial \xi =-(M/2)G_{ef}^{2}(\xi )
\end{equation}%
so that to leading order for $|C|<T_K$ %
\begin{equation}
S_{ef}(C)\simeq (|C|/T_{K})^{\Delta },\Delta =1/\gamma =G^{\ast }
\label{pow}
\end{equation}%
It should be noted that 
in the non-singular DOS case the power-law critical behavior like (\ref{pow}%
) takes place in a wide region, including $|C|>T_{K}$ and $|C|<T_{K}$ \cite%
{Gan}. Thus, unlike the situation of total screening (where the
strong-coupling region cannot be described by simple methods) we have an
interpolation description.

On taking into account higher orders in $1/M$ one has in the flat-band
one-impurity case \cite{Gan}
\begin{equation}
\Delta =\frac{2}{M}\left( 1-\frac{2}{M}\right) \simeq \frac{2}{M+2},
\label{Delts}
\end{equation}%
which agrees with the Bethe ansatz solution, see Ref. \cite{Cox}.








\section{Scaling behavior in the Kondo lattice}

As discussed above, the structure of perturbation theory is similar to
the non-singular case with the replacement $g_{ef}(\xi )\rightarrow
G_{ef}(\xi )$.
Now we pass to the lattice case, main point being inclusion of spin dynamics into scaling equations.
To establish properly the correspondence with the one-impurity case (\ref%
{Delts}), we may put $\gamma =M/2+1=1/\Delta .$ This yields at $M>2$
correct critical exponents for magnetic susceptibility, specific heat and
resistivity. The important case $M=2$ is more difficult from the theoretical
point of view: additional logarithmic factors occur in electronic specific heat and magnetic susceptibility, although the resistivity is still described by the $1/M$
expansion, see \cite{Cox,Col}.

We present below numerical results for $M=1$ ($\gamma =3/2$) and for $M=3$ ($%
\gamma =5/2$); the latter case may be relevant for Ce$^{3+}$ ion \cite{Cox}.

\subsection{Paramagnetic case}

The dependences $G_{ef}(\xi )$ and $\psi (\xi )$ from solution of the full scaling equations (\ref{sc2}) and (\ref{sc3}) in the paramagnetic phase are shown in Fig.1. During the
scaling process $\psi (\xi )$ increases according to (\ref{sc3}). One can
see that $G_{ef}(\xi )$ demonstrates a plateau at $G^{\ast }\simeq 1/\gamma $%
, which can be named a \textquotedblleft quasi-fixed point\textquotedblright.

Provided that the bare coupling parameter $g$ is not too small, at intermediate $\xi $ (near the plateau) we can put for rough estimations $G_{ef}^{{}}(\xi )\simeq G^{\ast
}=1/\gamma $ to obtain
\begin{equation}
\psi (\xi )\simeq a\gamma G_{ef}^{2}(\xi )\xi -a/\gamma g\simeq (a/\gamma
)(\xi -1/g)  \label{chi}
\end{equation}%
($\Psi (\xi >1)\simeq 1$). Thus a power-law behavior occurs
\begin{eqnarray}
\overline{\omega }_{ef}(C) &\simeq &\overline{\omega }(|C|/T_{K})^{\beta
},\quad \overline{S}_{ef}(C)\simeq (|C|/T_{K})^{\Delta },  \nonumber \\
~\beta &=&a/\gamma =a\Delta ,  \label{pm}
\end{eqnarray}%
which corresponds to the one-impurity NFL behavior (\ref{pow}).

The dependence (\ref{chi}) takes place up to the point
\begin{equation}
\xi _{1}\simeq (\lambda -\beta /g)/(1-\beta ).  \label{bound}
\end{equation}%
For $\xi >\xi _{1},$~$\psi (\xi )\simeq \psi (\xi _{1})\simeq \lambda \beta
/(1-\beta )$ is practically constant since $\Psi (\lambda +\chi -\xi )$
becomes small.




\begin{figure}[tbp]
\includegraphics[width=3.3in, angle=0,clip]{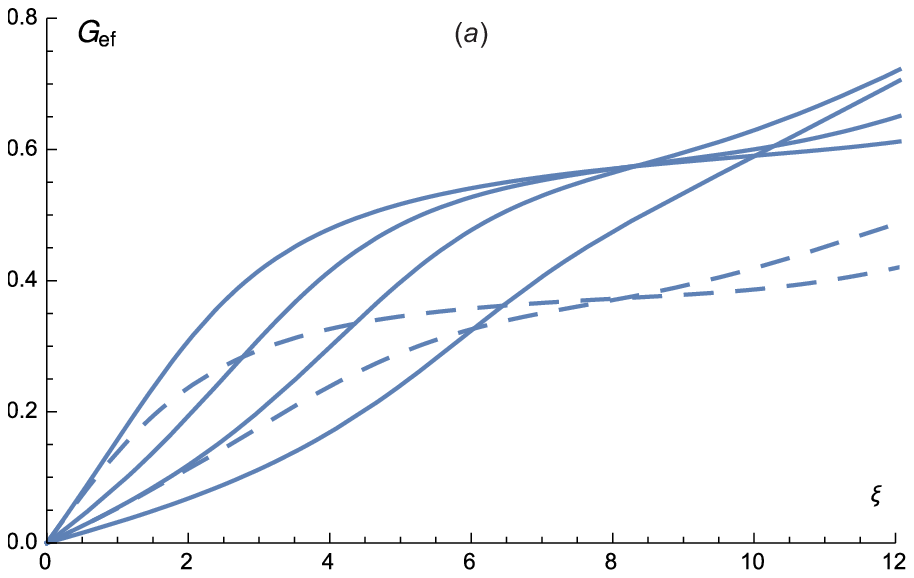}
\includegraphics[width=3.3in, angle=0,clip]{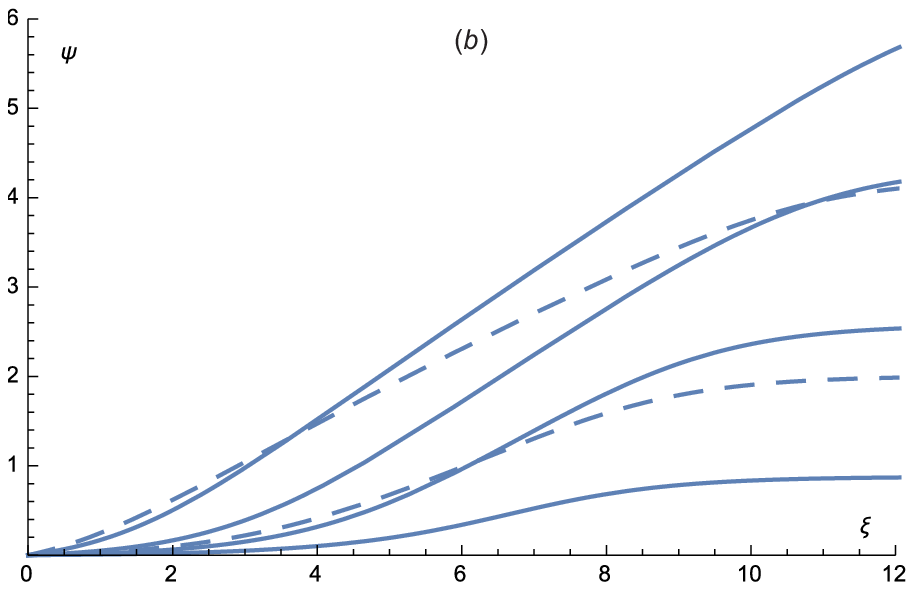}
\caption{The scaling trajectories for a paramagnet, $G_{ef}(\protect\xi )$
(a) and $\protect\psi (\protect\xi )$ (b). The parameter values are $\protect%
\lambda =6$, $a=0.7$, $g=0.03,0.05,0.08,0.15$, $M=1$  ($\protect\gamma = 3/2$, solid
lines) and $g=0.05,0.15$, $M=3$  ($\protect\gamma = 5/2$, dashed lines) -- for the
curves from below to above (when considering the left-hand part of the
figure) respectively }
\label{fig:1}
\end{figure}

\begin{figure}[tbp]
\includegraphics[width=3.3in, angle=0,clip]{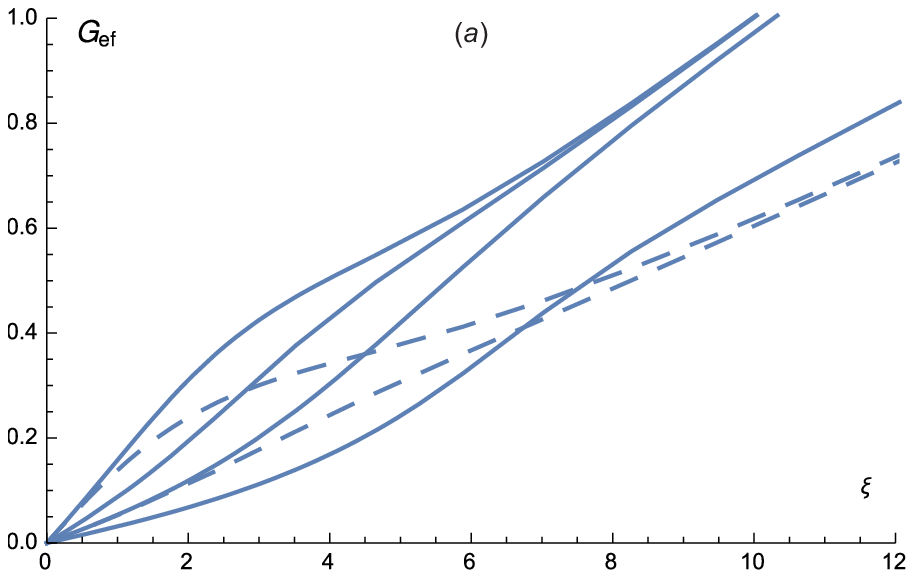} %
\includegraphics[width=3.3in, angle=0,clip]{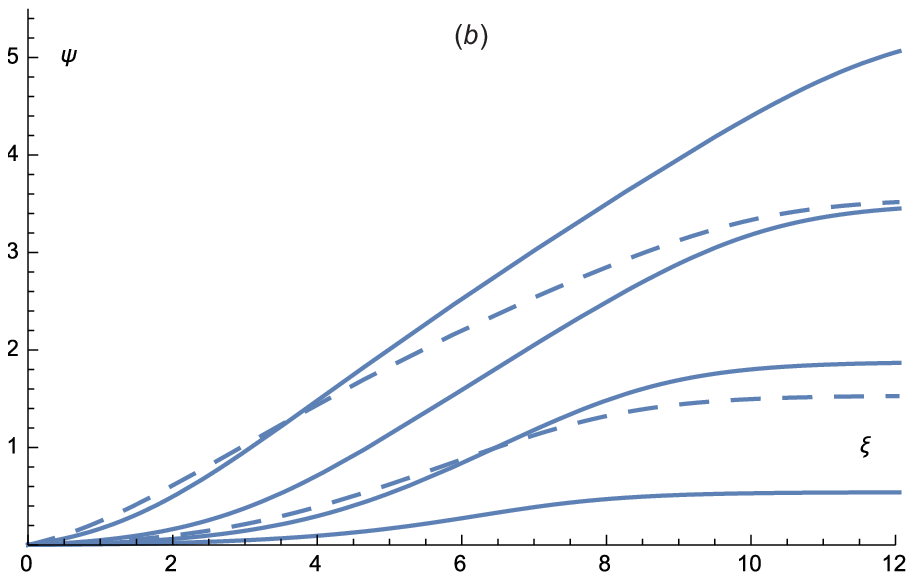}
\caption{The scaling trajectories for a paramagnet, $G_{ef}(\protect\xi )$
(a) and $\protect\psi (\protect\xi )$ (b). The singularity is shifted from
the Fermi level by $v=0.005$. Other parameter values are the same as in
Fig.1 }
\label{fig:2}
\end{figure}

When the DOS singularity is shifted from the Fermi level by the distance $v$%
, its influence on the scaling behavior becomes weaker. A similar effect
occurs when the logarithmic singularity is smeared (e.g., small electron
damping is introduced, $\ln E\rightarrow (1/2)\ln (E^{2}+\Gamma ^{2})$). At
small $v$ the scaling behavior is determined by a combined action of the
Kondo and Van Hove singularities. The influence of the shift is described by
the replacement $C\rightarrow C-v$ in the singular factors in the scaling
equations. The scaling trajectories for $v=0.005$ are shown in Fig.2.
The influence of the shift on the dependence $G_{ef}(\xi )$ is more
pronounced than that on $\psi (\xi )$ (for the latter, we have some
quantitative difference only).

On the other hand, at not too small $v$ and $g$ the DOS singularity becomes
unimportant, so that with increasing $\xi $ we come rapidly to a
non-physical fixed point, $g_{ef}(\xi )\rightarrow g^{\ast }$ with large $%
g^{\ast }$.

Since exact divergence of DOS is not required, we can suppose that not only
strong logarith mic singularities, but also weaker VHS (e.g., those in 3D
cubic lattices) may change considerably the scaling behavior. Note that the
increase of $\psi (\xi )$ in paramagnetic phase is much more stronger than
for a smooth DOS (cf. Ref. \cite{I16}).

The interval of bare coupling constant, where a NFL-like behavior occurs, is
very wide: we come to a quasi-fixed point independently of $g$, although
this point becomes unstable with increasing $\xi $. On the other hand, the
one-loop scaling yields for finite $M$ the NFL behavior in a narrow interval
of the bare coupling constant $g$ only, since with increasing $g$ we come
rapidly to strong-coupling regime where $g_{ef}(\xi >\lambda )\rightarrow
\infty ,$ the critical value $g_{c}$ being rather small \cite{I11}.
On the contrary, in the two-loop scaling there is no such a critical $g$
value at all: $g_{ef}(\xi )$ remains finite for any $g$ in the paramagnetic
case. Thus the lowest-order scaling cannot describe properly the case of not
too small $g$. 

In the region of the plateau ($G_{ef}(\xi )=$ const), $g_{ef}(\xi )$
decreases with increasing $\xi $. Therefore the scaling curves $G_{ef}(\xi )$
can intersect each other for different $g.$ However, this feature disappears
when introducing small shift of the singularity from the Fermi level, the
size of the plateau decreasing (Fig. 2).

From (\ref{pm}) we obtain the power-law dependence of magnetic
susceptibility
\[
\chi (T)\varpropto S_{ef}^{2}(T)/T\varpropto (T/T_{K})^{2\Delta -1}
\]%
%
%
%
As demonstrate NRG calculations for the one-impurity Kondo model with VHS  \cite{Zhur2016}, the local magnetic susceptibility

\begin{equation}
\chi _{\mathrm{loc}}(T)=\int\limits_{0}^{1/T} d\tau \langle S_{z}(\tau
)S_{z}\rangle  \   \label{ChiLocal}
\end{equation}%
(which just determines spin correlation functions and therefore corresponds
to present calculations) has a slight maximum tending to a constant value
with lowering temperature for $M=1$ and demonstrates a power-law NFL
behavior for $M=2$ (unlike logarithmic behavior in the flat-band case). Thus the tendency to NFL behavior increases in the presence of VHS.




\subsection{Ferromagnetic case}

Now we come to the situation of magnetic ordering. Of especial interest is
ferromagnetic state: its realization is connected with large density of
states at the Fermi level, so that peculiarities of the NFL state in the
presence of VHS should be also treated.

In magnetically ordered phases, the behavior for $\xi <\xi _{1}$ is similar
to that in a paramagnet, but the situation for $\xi >\xi _{1}$ changes since
the Van Hove singularity of $\Psi (\xi )$ at $\xi =0$ plays an important
role. Instead of decreasing, $\Psi (\lambda +\psi -\xi )$ starts to increase
at approaching $\xi _{1}$.

\begin{figure}[tbp]
\includegraphics[width=3.3in, angle=0]{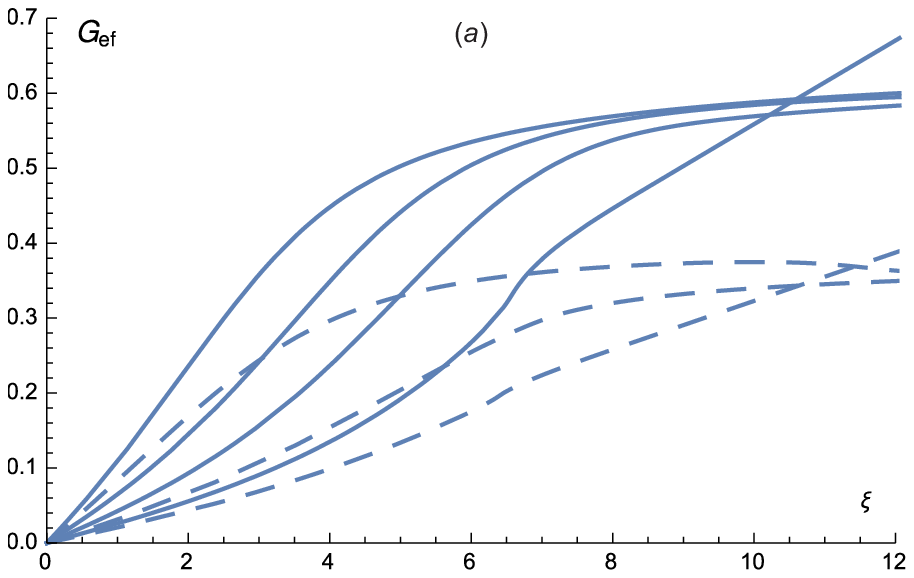} %
\includegraphics[width=3.3in, angle=0]{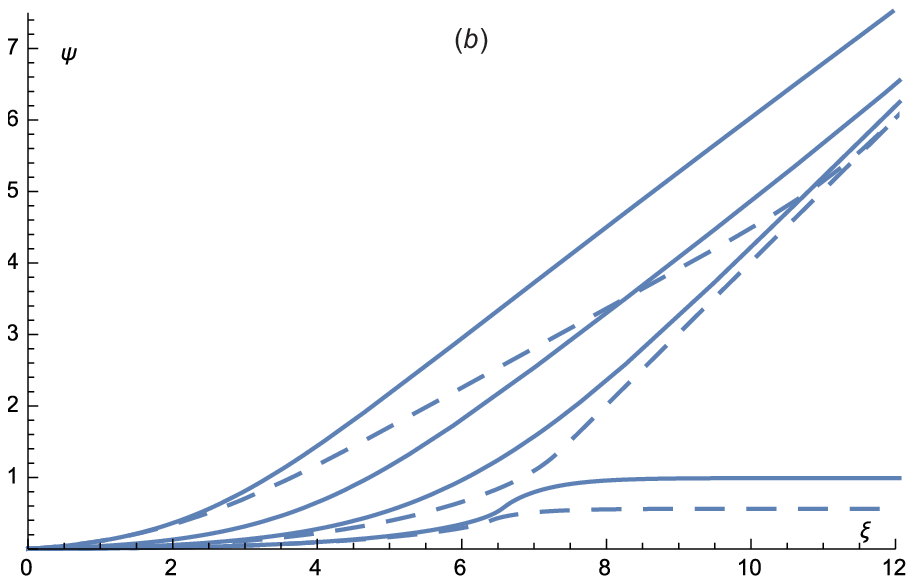}
\caption{ The scaling trajectories for a ferromagnet, $G_{ef}(\protect\xi )$
(a) and $\protect\psi (\protect\xi )$ (b). The parameter values are $\protect%
\lambda =6$, $a=1$, $g=0.025,0.04,0.06,0.1$, $M=1$  ($\protect\gamma = 3/2$, solid
lines) and $g=0.02, 0.03, 0.08$, $M=3$ ($\protect\gamma = 5/2$, dashed lines) -- for the
curves from below to above respectively, the damping parameter is $\protect%
\delta = 10^{-2}$ }
\label{fig:3}
\end{figure}

For sufficiently large $g$ (needed to reach the appreciable $ G_{ef}$ value during the
scaling increase at small $\xi $), provided that
\begin{equation}
a\gamma G_{ef}^{2}(\xi \simeq \xi _{1})\Psi ^{\max }\simeq a\gamma g^{\ast
2}\Psi ^{\max }\simeq (a/\gamma )\Psi ^{\max }>1,
\end{equation}%
at $\xi >\xi _{1}$ the argument of the function $\Psi $ in (\ref{sc3})
becomes almost constant (fixed), $\psi (\xi )\simeq \xi -\lambda $. Thus
further behavior is determined by the singularity of the scaling function
and is similar to that for smooth DOS \cite{I16}. We have for the frequency
and magnetic moment
\begin{equation}
\;\overline{\omega }_{ef}(C)\simeq |C|,~~\overline{S}_{ef}(C)/S\simeq (|C|/%
\overline{\omega })^{1/a}.  \label{lin}
\end{equation}


The scaling curves for a ferromagnet are shown in Fig.3 (In the
antiferromagnetic case the picture is qualitatively the same, cf. Ref. \cite%
{I16}).
\begin{figure}[tbp]
\includegraphics[width=3.3in, angle=0]{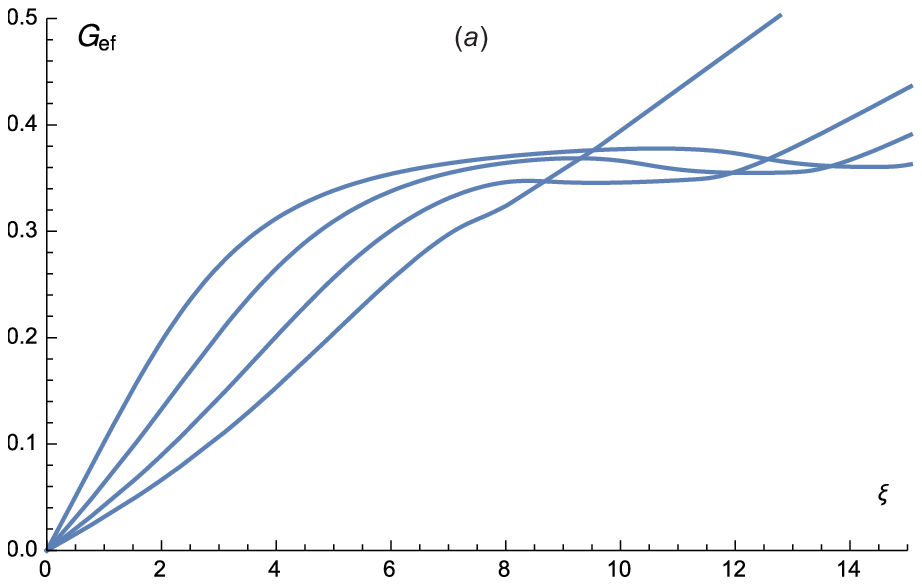} %
\includegraphics[width=3.3in, angle=0]{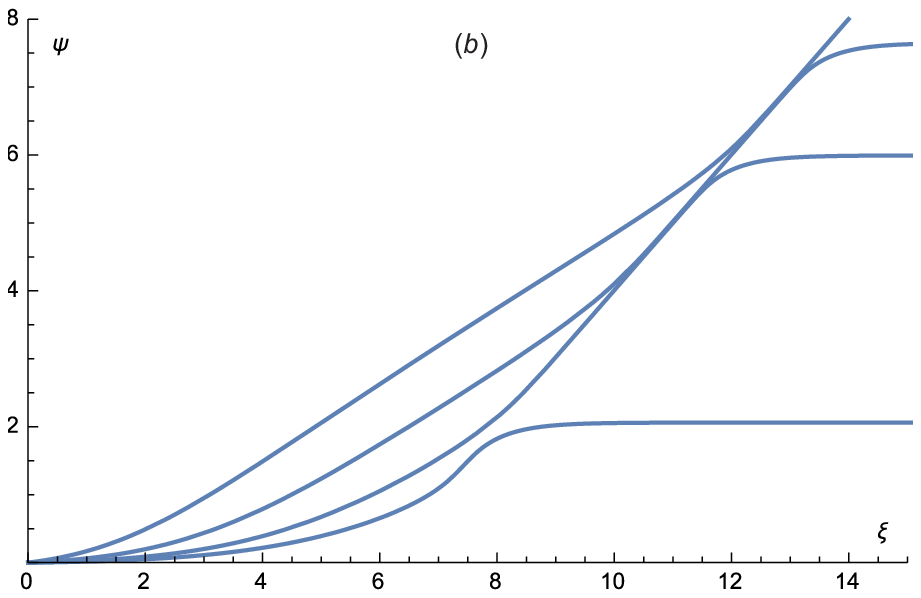}
\caption{ The scaling trajectories for a ferromagnet, $G_{ef}(\protect\xi )$
(a) and $\protect\psi (\protect\xi )$ (b). The parameter values are $M=3$ ($%
\protect\gamma = 5/2$), $\protect\lambda =6$, $a=1$, $g=0.03,0.04,0.06,0.1$
for the curves from below to above respectively, the damping parameter is $%
\protect\delta = 0.02$}
\label{fig:4}
\end{figure}

Thus the scaling behavior is changed at some critical value $g_{c}$. Above
the critical value $g_{c}$, the qualitative picture of the nearly linear scaling
trajectories $\psi (\xi )$ does not depend on $g$: they are almost parallel and slowly come together. 

In turn, there is a critical value of the damping parameter $\delta $ in the scaling  function (\ref{damp})  (which ia a
cutoff for the singularity determining $\Psi ^{\max
} $), so that for $\delta >\delta _{c}$ the infinite linear behavior does
not occur. This value, $\delta _{c},$ is determined by the values of $a$ and
$M.$
For large $\delta $ and $M>1$ the linear dependence of the type (\ref{lin})
can take place in a restricted region changing the behavior of the type (\ref%
{chi}), so that two NFL-like regions are observed (see Fig.4, $\delta _{c}$
is about 0.015 for $M=3$).

For $M=1$ the critical damping is not small: $\delta _{c}$ is about
0.1. This is favorable for occurrence of the NFL regime (\ref{lin}): it can
take place in the presence of a somewhat pronounced (even not too sharp) peak in
the scaling function $\eta $.
Of course, the simple model with constant damping can be generalized. So,
the increase of the damping with growing of $g_{ef}(\xi )$ was considered in
the scaling versions of Refs.\cite{731,I11}. As well as in the paramagnetic
case, a small shift of the singularity from the Fermi level results in a
change of $G_{ef}(\xi )$ behavior at large $\xi ,$ but influences weakly the
behavior $\psi (\xi )$.


Remember again that the quantity $\psi (\xi )$ determines the temperature
dependences of magnetic moment and thermodynamic characteristics (for the
corresponding discussion in the antiferromagnetic case, see Refs. \cite%
{731,I16}).


\section{Conclusions}

We have shown that the Kondo lattice with Van Hove singularities in electron spectrum demonstrates non-Fermi-liquid behavior in a wide parameter region.
As follows from (\ref{perturb}), the value of the Kondo
temperature is rather high.
Thus the system is characterized by moderate specific heat, but large magnetic
susceptibility.  The
renormalization of magnetic
moment  is much stronger than in the case of smooth DOS.
Although the heavy-fermion behavior is not
expressed, a tendency to magnetic ordering occurs, which is
characteristic  for weak itinerant magnets with VHS like ZrZn$_{2}$ too.

In this connection, we can also mention some experimental examples of NFL features for ruthenate d-systems.
An enhancement of the electronic specific heat and   magnetic susceptibility
was observed in the layered  system Sr$_{2-x}$La$_x$RuO$_4$ with increasing x, the Fermi-liquid behavior being violated near the critical value x = 0.2. Such a tendency is explained by the elevation of the Fermi energy toward VHS of the thermodynamically dominant Fermi-surface sheet. The NFL behavior is attributed to two-dimensional ferromagnetic fluctuations with short-range correlations at VHS \cite{rut}.
The bilayered ruthenate  Sr$_3$Ru$_2$O$_7$ is a paramagnetic  Fermi liquid with strongly enhanced quasiparticle masses \cite{rut1}. The Fermi-liquid region of the phase diagram is  suppressed by magnetic field, and NFL behavior extends up to very low temperatures upon approaching the critical metamagnetic field $B$ = 7.8 T \cite{rut2}.

Occurrence of giant Van Hove singularities (which
are important, e.g., for ferromagnetism of iron) is intimately connected with
intersection of more weak singularities, i.e. with degeneracy of electron
bands \cite{Vons}. Already in the classical textbook on magnetism \cite%
{Mattis} such a degeneracy is considered as a key to itinerant
ferromagnetism. This statement is in spirit of a multichannel model too.

The results obtained for the scaling behavior are qualitatively reliable
for $M>2$. On the other hand, they, generally speaking, should be verified by more strict analytical and numerical
methods, including two-loop field-theoretical or functional renormalization group (fRG) (see, e,g., investigations in Refs. \cite{Kat,Kat1,Eb} performed  for the Hubbard model). It should be noted that a scaling treatment in the presence of logarithmic singularities meets with difficulties \cite{Yakovenko}. The corresponding
problems of higher-order scaling are also discussed in recent works \cite%
{Kapustin,Kapustin1}.
Probably, the nesting problem (see, e.g., \cite{Dz}) can be considered in a similar way,
but strong \textbf{k}-dependence requires a more sophisticated treatment
without averaging over the Fermi surface.
The case of the Kondo lattice seems to be more simple since the scaling equations are obtained from those for the one-impurity model by inclusion of spin dynamics.

The author is grateful to A.A. Katanin for useful discussions.
The research was carried out within the state assignment of FASO of Russia
(theme ``Quantum'' No. 01201463332). This work was supported in part by
Ural Branch of Russian Academy of Science (project no. 15-8-2-9) and by the
Russian Foundation for Basic Research (project no. 16-02-00995).

\end{document}